\documentclass[10pt, conference, letterpaper]{IEEEtran}
 
\IEEEoverridecommandlockouts
\usepackage{cite}
\usepackage{amsmath,amssymb,amsfonts}
\usepackage{algorithm}
\usepackage[noend]{algpseudocode}
\usepackage{graphicx}
\usepackage{textcomp}
\usepackage[table, dvipsnames]{xcolor}
\usepackage{svg}
\svgpath{{images}}
\usepackage{stfloats}
\usepackage{pdfpages}
\usepackage{tabularx}
\usepackage{subcaption}
\usepackage[export]{adjustbox}
\usepackage{tikz}
\usepackage{booktabs}
\usepackage{todonotes}
\setuptodonotes{inline}

\usetikzlibrary{decorations.pathreplacing}
\usepackage{pifont} 

\newcommand{\cmark}{\ding{51}}
\newcommand{\xmark}{\ding{55}}

\usepackage{caption}

\def\BibTeX{{\rm B\kern-.05em{\sc i\kern-.025em b}\kern-.08em
    T\kern-.1667em\lower.7ex\hbox{E}\kern-.125emX}}

\makeatletter
\def\BState{\State\hskip-\ALG@thistlm}
\makeatother

\begin{document}
\title{Emulation vs Simulation: A Case Study from Congestion Control Algorithms in Low Earth Orbit Satellite Networks}

\author{
\IEEEauthorblockN{Aiden Valentine$^{*}$, Mihai Mazilu$^{*}$, James Knowles$^{*}$, Ian Wakeman$^{*\dagger}$  and George Parisis$^{*}$}
\IEEEauthorblockA{School of Engineering and Informatics, University of Sussex$^{*}$\\ Zhejiang Gongshang University$^{\dagger}$\\
\{a.valentine, m.mazilu, ck484, ianw, g.parisis\}@sussex.ac.uk}
}

\maketitle

\begin{abstract}

Evaluating congestion control is inherently challenging because performance depends on the interaction between the congestion-control algorithm, transport stack, application behaviour, measurement process, and network dynamics. This challenge is growing as state-of-the-art protocols incorporate pacing, selective loss recovery, model-based control, and, more recently, reinforcement learning. Low Earth Orbit (LEO) satellite networks are a particularly demanding setting: rapidly changing paths, handovers, non-congestive loss, RTT variation, and transient hotspots all affect transport behaviour. This paper reports the lessons learned from an extensive evaluation campaign across both simulation and emulation for LEO satellite congestion control. We compare multiple classes of congestion-control algorithms, including Cubic, BBR variants, LEO-specific protocols, and reinforcement-learning-based control, using comparable implementations across OMNeT++/INET simulation and Mininet-based emulation with the Linux transport stack. This gives us a rare opportunity to examine not only protocol performance, but also the methodological strengths and limitations of each experimental environment. Our findings show that simulation is indispensable for constellation-scale exploration, controlled parameter sweeps, and future deployment studies, but can miss behaviours caused by production transport-stack mechanisms such as pacing, SACK, RACK, kernel timing, and rate sampling. Emulation exposes these implementation-dependent effects and provides a necessary validation step, but is harder to scale and less exactly repeatable. We distil these experiences into practical lessons for combining simulation and emulation to obtain results that are scalable, reproducible, and deployment-relevant.

\end{abstract}

\section{Introduction}
\label{sect:introduction}

The study of congestion control (CC) remains central to the design and operation of modern communication systems. As applications demand higher throughput, lower latency, and more robust performance across heterogeneous paths, transport protocols such as Cubic~\cite{cubic}, BBR~\cite{bbrv1}, and emerging alternatives continue to evolve. Evaluating these protocols is inherently difficult; performance depends not only on the CC algorithm, but also on the surrounding transport stack, application behaviour, measurement process, and network dynamics. This difficulty is increasing as state-of-the-art protocols rely on pacing, selective loss recovery, model-based estimation, and reinforcement learning, all of which are sensitive to implementation details and experimental methodology. Controlled experimental environments are therefore essential. Simulation and emulation are the two dominant approaches. Simulation, exemplified by discrete-event frameworks such as OMNeT++~\cite{omnetpp_release_2026}, provides scalability, reproducibility, and fine-grained control over topology, traffic, and protocol internals. Emulation, using tools such as Mininet~\cite{mininet}, executes real protocol stacks and applications under controlled network conditions, exposing implementation-level effects that are often absent from simulation. However, the relative strengths and limitations of these approaches are difficult to assess in isolation, particularly when the target network is highly dynamic and the protocols under test depend on complex transport-stack behaviour.

Low Earth Orbit (LEO) satellite networks provide a demanding and increasingly important setting for this comparison. LEO paths experience frequent handovers, route changes, RTT variation, non-congestive loss, and transient congestion hotspots. These dynamics challenge conventional CC assumptions and have motivated LEO-aware protocols such as SaTCP~\cite{satcp} and LeoCC~\cite{leocc}. Recent work, including our experimental study of CC over LEO satellite networks~\cite{mazilu2026leo}, has shown that these dynamics can substantially alter protocol behaviour under real-stack execution. However, existing evaluations typically focus on either simulation or emulation, leaving limited guidance on how constellation-scale simulation and real-stack emulation should be combined for LEO CC research.

This paper brings together a set of complementary technologies to enable a direct comparison of simulation and emulation for LEO CC evaluation. On the simulation side, we build on our OMNeT++/INET LEO constellation modelling~\cite{valentine_developing_2021} and extended TCP implementations. On the emulation side, we use Mininet-based LEO path emulation, executing CC protocols through the Linux transport stack. We also incorporate RayNet, our reinforcement-learning (RL) system for CC experimentation \cite{giacomoni2023raynetsimulationplatformdeveloping}. Combined with state-of-the-art protocols, including Cubic, BBR variants, SaTCP, LeoCC, and Orca, this creates a common evaluation workflow spanning constellation-scale simulation, real-stack emulation, and learning-based control.

Rather than treating simulation and emulation as interchangeable evaluation tools, we use this workflow to examine what each environment makes visible and what each abstracts away. Across extensive experiments, we identify cases where the two methodologies agree, cases where they diverge, and the modelling or implementation details that explain the gap. These include missing TCP mechanisms in simulators, difficulties in modelling LEO handovers and routing, differences in metric collection, challenges in reproducing Linux transport-stack behaviour, and additional complications introduced by RL-based CC. The result is a set of practical lessons for using simulation and emulation together, rather than relying on either methodology in isolation.

The remainder of the paper first establishes the evaluation landscape, including the LEO dynamics, protocol classes, and experimental tools relevant to this work. We then describe the simulation and emulation methodology used to obtain comparable results, before presenting the lessons learned from these experiments. We conclude with guidance on how simulation and emulation can be combined for scalable and deployment-relevant LEO CC evaluation.

\section{Evaluation Context}
\label{sect:technology}

This section identifies the technical landscape behind our evaluation by explaining which LEO dynamics, congestion-control protocols, and experimental tools matter when results are compared across simulation and emulation. We first identify the LEO network dynamics that make CC evaluation difficult, then relate those dynamics to the protocol classes considered in this paper: loss-based, model-based, LEO-aware, and learning-based CC. This framing exposes three concrete requirements. First, any LEO CC evaluation requires path dynamics that reflect the behaviours seen by transport flows, including changes in connectivity, routing, delay, capacity, and loss. Second, CC evaluation requires transport-stack mechanisms that are unevenly supported in simulation frameworks, including pacing, SACK, RACK, delivery-rate estimation, and Linux-style loss recovery. Third, learning-based CC adds another layer of complexity; the evaluation environment must expose observations, actions, rewards, and timing to an external learning system at scale. These requirements motivate the methodology and evaluation synthesis attempted in this paper over LEO evaluation components spanning our constellation model and Mininet-based emulation, extended OMNeT++/INET transport support, and RayNet for learning-based CC experimentation.

We begin with the LEO setting, since its path dynamics define the evaluation problem. LEO satellite networks deploy large constellations of satellites that move rapidly, around $7.5$ km/s, at much lower altitudes than geostationary systems. This reduces propagation delay, but creates short visibility windows and makes connectivity highly dynamic. The resulting coverage and path behaviour are determined largely by orbital altitude and inclination. For example, the first Starlink shell operates at an altitude of 550 km and an inclination of 53$^{\circ}$. Higher-inclination shells extend coverage towards polar regions, whereas lower-inclination shells concentrate capacity closer to the equator. These orbital properties require frequent changes in connectivity between satellites, user terminals, and ground infrastructure~\cite{kassing_exploring_2020, cao_satcp_2023}.

\begin{figure}[h]
    \centering
    \includegraphics[width = 250pt]{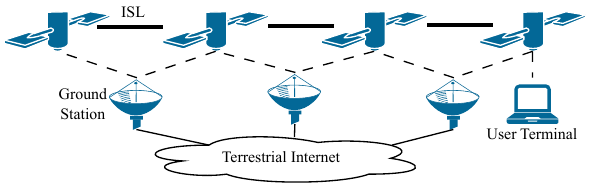}
    \caption{An example of a LEO satellite network structure.}
    \label{fig:leoExample}
\end{figure}

LEO satellite communication depends on three main infrastructure components: user terminals (UTs), ground stations (GSs), and, increasingly, inter-satellite links (ISLs). UTs provide end-user connectivity, while GSs act as gateways between the satellite constellation and the terrestrial Internet. GSs can use multiple phased-array antennas to maintain simultaneous connections to several satellites~\cite{wong_network_2024}. UTs are typically more constrained and may maintain only one active satellite connection at a time. Traffic can be routed either through bent-pipe (BP) forwarding, where satellites relay packets between UTs and GSs, or through ISL-enabled paths, where packets are forwarded across multiple satellites before reaching a ground station. BP paths can be efficient over shorter distances, especially when endpoints are close to the relevant satellite and gateway~\cite{handley_using_2019}. ISL-enabled paths can reduce RTT and improve throughput for some long-distance connections~\cite{hauri_internet_2020}. Figure~\ref{fig:leoExample} illustrates this network structure.

These architectural properties create transport-layer challenges that are central to this paper. Because satellites move continuously, the end-to-end path can change through handovers, route updates, changing ISL availability, or changing GS reachability. The resulting effects include non-congestive latency variation and loss, frequent handover disruption, and transient congestion hotspots. These dynamics complicate CC evaluation because transport protocols may interpret mobility-induced loss or RTT changes as congestion. They also make the choice of experimental environment important; simulation is needed to explore constellation-scale path dynamics, while emulation is needed to validate how real transport stacks respond to selected LEO-like events.

We therefore select protocols that exercise different assumptions about network feedback and different requirements from the experimental environment. Cubic represents loss-based control, where packet loss is treated as the main congestion signal~\cite{cubic}. BBRv1 and BBRv3 represent model-based control, where the sender estimates bottleneck bandwidth and propagation RTT and relies heavily on pacing, delivery-rate estimation, ACK processing, and loss recovery~\cite{bbrv1,bbrv3}. SaTCP and LeoCC are LEO-aware adaptations: SaTCP extends Cubic by using predicted handover or route-update events to avoid unnecessary congestion-window reductions~\cite{satcp}, while LeoCC builds on BBRv1 by detecting LEO reconfiguration events and refreshing its bandwidth and RTT model when the path changes~\cite{leocc}. Orca represents learning-based CC, where behaviour depends not only on transport-stack mechanisms, but also on the timing and granularity of observations and control actions~\cite{orca}. Learning-based CC also introduces an additional tooling requirement not captured by Table~\ref{tab:leo-tools}; the simulator must exchange observations, actions, and rewards with an external learning framework at scale. RayNet addresses this requirement by exposing OMNeT++ simulations as RL environments~\cite{giacomoni2023raynetsimulationplatformdeveloping}, enabling RL-based CC experiments to be integrated into the same simulation workflow. Together, these protocols form a representative set for exposing the simulation--emulation gap. Table~\ref{tab:tcp-features} summarises native support for these protocols and the transport mechanisms they depend on across common simulation and emulation environments. Section \ref{sect:methodology} describes the extensions we implemented to close these gaps.

\begin{table}[h]
\centering
\setlength{\tabcolsep}{15pt}
\begin{tabular}{lccc}
\hline
 & \multicolumn{2}{c}{Simulation} & Emulation \\
 & OMNeT++/INET & ns-3 & Mininet \\
\hline
\multicolumn{4}{l}{\textbf{TCP Variants}} \\
Cubic    & \xmark & \cmark & \cmark \\
BBRv1    & \xmark & \cmark & \cmark \\
BBRv3    & \xmark & \xmark & \cmark \\
SaTCP    & \xmark & \xmark & \cmark \\
LeoCC    & \xmark & \xmark & \cmark \\
Orca     & \xmark & \xmark & \cmark \\
\hline
\multicolumn{4}{l}{\textbf{TCP Mechanisms}} \\
Pacing   & \xmark & \cmark & \cmark \\
SACK     & \cmark & \cmark & \cmark \\
RACK     & \xmark & \xmark & \cmark \\
\hline
\end{tabular}
\caption{Availability of TCP variants and mechanisms across simulators and emulation, before our extensions.}
\label{tab:tcp-features}
\end{table}

The available LEO tooling reflects the same split, as summarised in Table~\ref{tab:leo-tools}. Constellation simulators can explore large-scale path dynamics and future deployments, but they do not execute the production transport stack. Emulators execute real stacks, but are harder to scale and often reproduce only selected LEO paths. Hypatia and our LEO model support constellation-scale simulation, making them suitable for exploring topology evolution, routing changes, and geographically diverse path dynamics~\cite{kassing_exploring_2020,valentine_developing_2021}. In contrast, LeoEM and LeoReplayer provide real-stack evaluation of selected or replayed LEO path dynamics, which is valuable for controlled validation but less suitable for arbitrary constellation-wide exploration~\cite{satcp,leocc}. StarryNet and Xeoverse target broader LEO emulation, but are limited by openness or scalability constraints~\cite{starrynet,xeoverse}.

\begin{table}[h]
\centering
\setlength{\tabcolsep}{5pt}
\begin{tabular}{lcccc}
\hline
Tool & Type & Open & Scope & Real stack \\
\hline
Our LEO model~\cite{valentine_developing_2021} & Simulation & \cmark & Constellation & \xmark \\
Hypatia~\cite{kassing_exploring_2020} & Simulation & \cmark & Constellation & \xmark \\
StarryNet~\cite{starrynet} & Emulation & Partial & Limited & \cmark \\
Xeoverse~\cite{xeoverse} & Emulation & \xmark & Constellation & \cmark \\
LeoEM~\cite{satcp} & Emulation & \cmark & Selected path & \cmark \\
LeoReplayer~\cite{leocc} & Emulation & \cmark & Trace replay & \cmark \\
\hline
\end{tabular}
\caption{LEO evaluation tooling relevant to this paper.}
\label{tab:leo-tools}
\end{table}

Together, Tables~\ref{tab:tcp-features} and~\ref{tab:leo-tools} highlight the core methodological issue: no single environment provides all of the required properties. Simulation gives scale, controllability, and integration with LEO constellation models and RL training, but common simulator TCP stacks lag behind production systems. Emulation gives real Linux transport behaviour, including mechanisms such as pacing, SACK, RACK, and kernel rate sampling, but is harder to scale and less exactly repeatable. The remainder of the paper examines how these environments can be combined, and where discrepancies remain even after extending the simulator to support the required transport mechanisms.

\section{Experimental Methodology}
\label{sect:methodology}
To allow for exploration of the design space of CC over LEO satellite networks, we opted to build around our existing expertise with OMNeT++ \cite{omnetpp_release_2026} and the INET framework \cite{inet_release_2026} for our simulations.  This is very typical of CC research, since simulation allows for the fine grained control over topology, traffic and protocol internals, allowing for exploration over both mechanism design and the parameter space.  We implemented Cubic \cite{cubic}, BBRv1 \cite{bbrv1}, BBRv3 \cite{bbrv3}, SaTCP \cite{satcp}, LeoCC \cite{leocc} and Orca \cite{orca} within OMNeT++/INET. We also updated INET’s TCP stack by refining the SACK implementation to better reflect the Linux kernel counterpart and added support for the RACK loss recovery mechanism \cite{cheng_rack-tlp_2021}. RACK is used by all protocols in our evaluation and reflects common practice in modern TCP variants. All source code has been made publicly available to support reproducibility and future research.\footnote{https://figshare.com/s/4cc748e3034b054f6448}

Our emulation environment is built using Mininet, with all network nodes represented as Linux network namespaces on a single host running Ubuntu 22.04 LTS with a custom BBRv3-enabled kernel. The host is equipped with an AMD Ryzen Threadripper PRO 7965WX 24-core processor and 64 GB of memory. We employed LeoEM \cite{satcp} to emulate real LEO satellite network paths. LeoEM is a lightweight, Mininet-based emulation framework that models user terminals, ground stations as well as hard and soft handovers. We emulate different base RTTs and non-congestive loss values by configuring the Linux traffic control (\textit{tc}) queuing discipline \textit{netem}; bandwidth shaping is done using \textit{tbf} with drop-tail FIFO queues. The source code for the frameworks and plotting as well as all data and individual experiments is available online\footnote{https://github.com/Aruuni/mininettestbed/tree/ifip2026}. We have set the socket buffers (both send and receive) to a high value to prevent bottlenecks. 

\subsection{Metric Collection}
In discrete event network simulation frameworks such as ns-3 and OMNeT++, simulation time proceeds through the events scheduled in their respective event queues. Thus, every variable can be observed at the exact moment in simulation time that it changes. OMNeT++ does so through an explicit emit signal which records the traced variable, causing no overhead to the simulation itself, as the underlying execution is not real-time. 

Conversely, emulation offers no equivalent mechanism. Therefore, tracing emulation variables must be done carefully: sampling too frequently can distort the measurements, while sampling too coarsely can hide the short-timescale behaviour that is important for CC analysis. To balance these constraints, we generate traffic using iPerf3, which reports throughput/goodput at a minimum interval of 100ms. This polling granularity inherently hides rapid transients, including bursts, sudden window changes, or fast retransmits. To combat this, we supplement iPerf3 with per-socket kernel statistics obtained via $ss$, which exposes internal TCP state and provides more granular 10ms reporting of values such as smoothed RTT and cwnd. The flexibility of simulation allows us to match the $ss$ interval in OMNeT++ by emitting traced variables at a similar interval.

\section{Lessons Learnt}
\begin{figure*}[!ht]
    \centering
    \includegraphics[width=\textwidth]{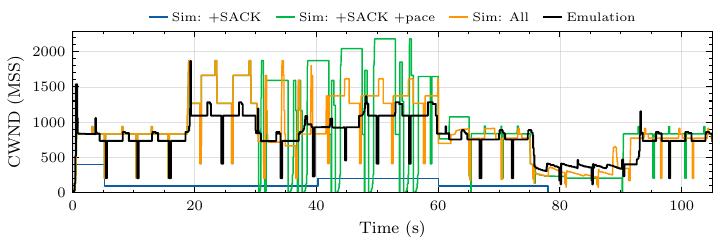}

    \caption{Incremental TCP mechanism comparison for BBRv3}
    \vspace{-0.4cm}
    \label{fig:ablation}
\end{figure*}

Our initial challenge in simulation was to develop an accurate representation of LEO satellite trajectories~\cite{valentine_developing_2021}. This required additional effort beyond importing satellite positions into OMNeT++. Substantial work was needed to ensure that satellites followed realistic orbital propagation and that this movement was translated into the network events that affect transport-layer behaviour. In particular, the model had to represent UT operation, GS deployment, inter-satellite connectivity, link updates, hard handovers, and end-to-end routing. Each of these mechanisms affects the path seen by a transport flow. A hard handover may introduce a short outage or packet loss; a route update may change the propagation delay; a new satellite-terminal association may change the bottleneck link; and global routing decisions determine whether these changes are localised or affect the full end-to-end path. This was a central difficulty in building the simulation framework because many of these behaviours are difficult to validate directly. Commercial LEO operators do not expose the internal details of their routing, handover, or terminal-association mechanisms~\cite{satcp}. As a result, even a detailed constellation simulator necessarily contains modelling assumptions. This is both a strength and a weakness: simulation allows us to explore future deployments and controlled scenarios, but the resulting CC behaviour may depend on assumptions that are not directly observable in deployed systems. 

This motivates the complementary use of emulation. The LEO simulation framework is best suited to generating constellation-scale path dynamics, including routing changes, handovers, and geographically dependent RTT variation. Emulation is then used to replay selected LEO-like dynamics under real transport-stack behaviour. The following results should not be read as a claim that either LEO model exactly reproduces a commercial constellation. Rather, they show how conclusions change as constellation-scale modelling and real-stack execution are combined. This distinction is important because commercial LEO routing, handover, and terminal-association policies are not directly observable, so both simulation and emulation necessarily depend on modelling choices. This is also why constellation-scale simulation remains important for future work on multiflow mechanisms and congestion hotspots that emerge across the wider constellation.

\begin{figure*}[!ht]
    \centering
    \includegraphics[width=\textwidth]{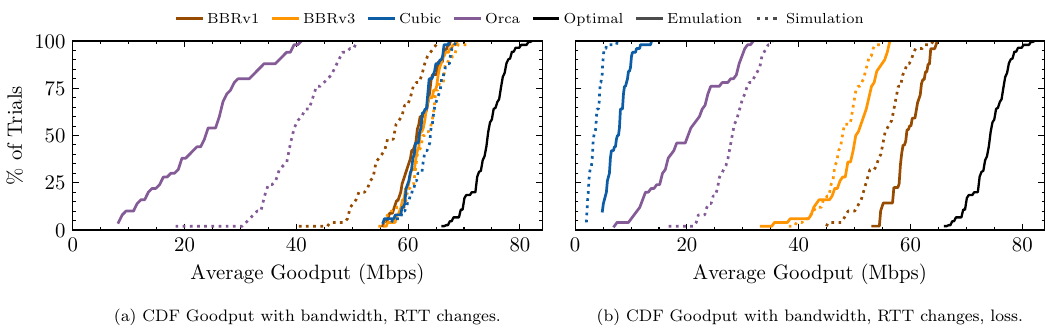}

    \caption{Responsiveness: Cumulative goodput distributions in dynamic networks.}
    \vspace{-0.4cm}
    \label{fig:responsiveness}
\end{figure*}


\subsection{Fidelity of Transport Stack Behaviour}
\label{ablation}
An ongoing difficulty with simulation frameworks is that their TCP stacks do not reflect the feature set available in production systems. The Linux kernel ships mature implementations of pacing, selective acknowledgement (SACK) \cite{floyd_tcp_1996}, recent acknowledgement (RACK) loss detection \cite{cheng_rack-tlp_2021}, and a broad range of CC algorithms, including Cubic, BBR, and BBRv3, all of which are actively deployed on the public Internet. OMNeT++'s INET stack, as of OMNeT++ 6.4.0 and INET 4.6.0, provides partial SACK support, but lacks pacing, RACK, and native implementations of Cubic, BBR, and BBRv3 \cite{omnetpp_release_2026,inet_release_2026,inet_tcp_docs_2026}. ns-3, as of version 3.45, is closer to production TCP behaviour: it supports pacing and includes Cubic and BBR. However, it does not provide RACK loss detection or BBRv3, so current loss-recovery behaviour and the latest model-based CC schemes cannot be reproduced directly \cite{ns3_release_2025,ns3_tcp_docs_2026}. These gaps are especially important in LEO networks, where reordering, short outages, and abrupt RTT shifts can make transport behaviour depend heavily on acknowledgement processing and loss recovery. We reduced this gap by extending the OMNeT++/INET TCP stack and contributing the changes back to the open-source repository. Building on INET's existing SACK functionality, we implemented SACK scoreboard behaviour, packet pacing, RACK loss detection, and forward acknowledgement (FACK)-style recovery\cite{rfc2018,rfc2883,rfc8985,mathis_fack_1996}. These extensions provided the transport mechanisms needed to implement and evaluate a wider set of CC models in OMNeT++/INET, including BBRv1, BBRv3, a paced variant of Cubic, and LEO-specific schemes such as SaTCP and LeoCC. 

Emulation was essential for validating these extensions. Matching the high-level CC logic is insufficient if the surrounding TCP mechanisms behave differently from a production stack. We therefore used Linux-based emulation as a reference point to check whether the extended INET implementation produced qualitatively similar behaviour under controlled LEO-like dynamics. This experience highlights a broader lesson: simulator fidelity requires tracking not only new CC algorithms, but also the supporting TCP mechanisms that shape their behaviour.

To illustrate the effect of these implementation details, we use a single-flow fidelity progression experiment comparing successive OMNeT++/INET implementations of BBRv3 against Linux-based emulation, shown in Figure \ref{fig:ablation}. The path begins with a 100Mbps bottleneck, 50ms RTT, and a 1 bandwidth-delay product (BDP) queue. At 15s, the bottleneck bandwidth is doubled; at 30s, it is restored to 100Mbps; at 35s, the RTT is doubled; at 60s, it is restored to 50ms; at 75s, a 1\% non-congestive loss rate is introduced; and at 90s, the loss is removed. We compare BBRv3 in emulation against three simulated variants: SACK only, SACK with pacing, and SACK with pacing and RACK. With only SACK, the simulated BBRv3 sender cannot pace traffic, which is fundamental to BBR's model of sending at the estimated bottleneck rate. Adding pacing improves correspondence with emulation, but the simulated sender still diverges once loss is introduced. Without RACK, loss detection remains tied more closely to duplicate acknowledgements and SACK, rather than the time-based loss inference used by the Linux implementation. Under rapid bandwidth and RTT changes as well as non-congestive loss, BBRv3 changes when it enters recovery, which packets it retransmits, and how quickly it restores its delivery-rate model. Adding RACK therefore brings the simulated congestion-window evolution closer to the emulated implementation. 

Even with SACK, pacing, and RACK enabled, the simulated and emulated traces do not match perfectly. This residual gap reflects implementation details that are difficult to reproduce in a discrete-event model, including ACK timing, send-buffer state, timer precision, and fixed-point arithmetic in BBR's internal calculations. The experiment therefore highlights the central difficulty of simulator fidelity: modern congestion control depends on the surrounding transport machinery as much as on the algorithm itself. Simulation can be made substantially more faithful, but emulation remains necessary to test whether those improvements replicate real-stack behaviour under rapidly changing LEO-like path conditions.

We follow this single-flow fidelity progression with a more dynamic responsiveness experiment, in which bandwidth, RTT, and loss change simultaneously and at a higher frequency. This tests whether the agreement observed in the mechanism-level comparison persists under more realistic LEO-like path dynamics.

We repeat a single flow 50 times, varying the bandwidth and RTT every 15 seconds (shown in Figure \ref{fig:responsiveness}a) and, in a second configuration, additionally injecting random loss (shown in Figure \ref{fig:responsiveness}b). We then replay the identical schedule of changes in simulation, allowing a direct comparison of how the two environments track the same rapidly varying conditions for a total of 300 seconds. 

The responsiveness observed in simulation is largely consistent with the corresponding emulation results, reinforcing the robustness of our implementation. In the case of Orca, the responsiveness is similarly poor, primarily due to the choice of states and the reward formulation. This agreement is particularly important in the loss configuration, where the simulated stack must rely on the newly added SACK scoreboard and RACK-based recovery logic rather than only on the CC schemes themselves. Since loss recovery was one of the main sources of divergence, the close correspondence under random loss suggests that these mechanisms now interact with the schemes (BBR variants in particular) in a way that is qualitatively aligned with the Linux reference implementation.

\subsection{Fairness analysis under LEO conditions}
\begin{figure*}[t]
    \centering

    \begin{subfigure}[t]{0.48\textwidth}
        \centering
        \includegraphics[width=\linewidth]{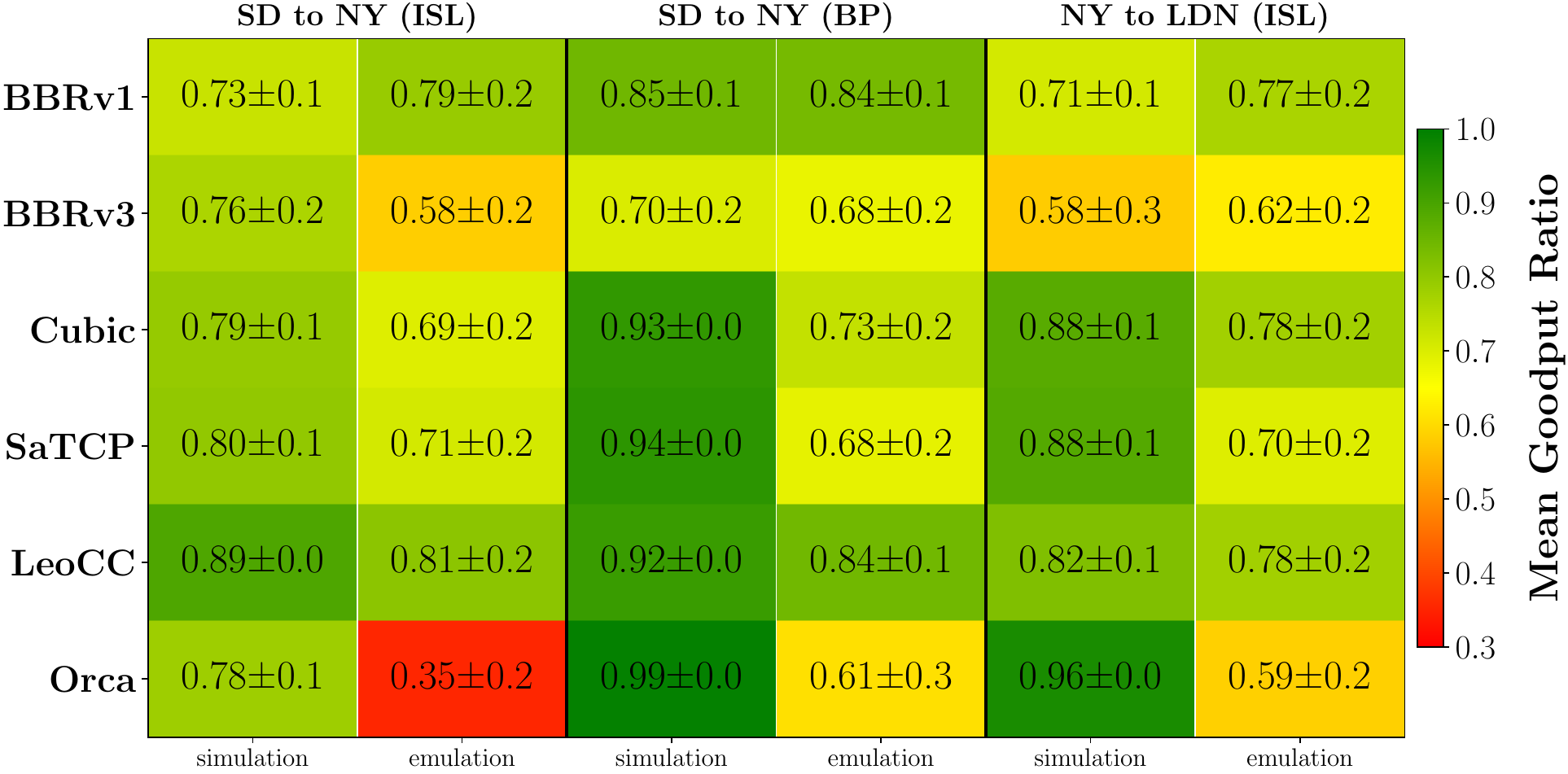}
        \caption{Fairness}
        \label{fig:fairness_heatmap}
    \end{subfigure}
    \hfill
    \begin{subfigure}[t]{0.48\textwidth}
        \centering
        \includegraphics[width=\linewidth]{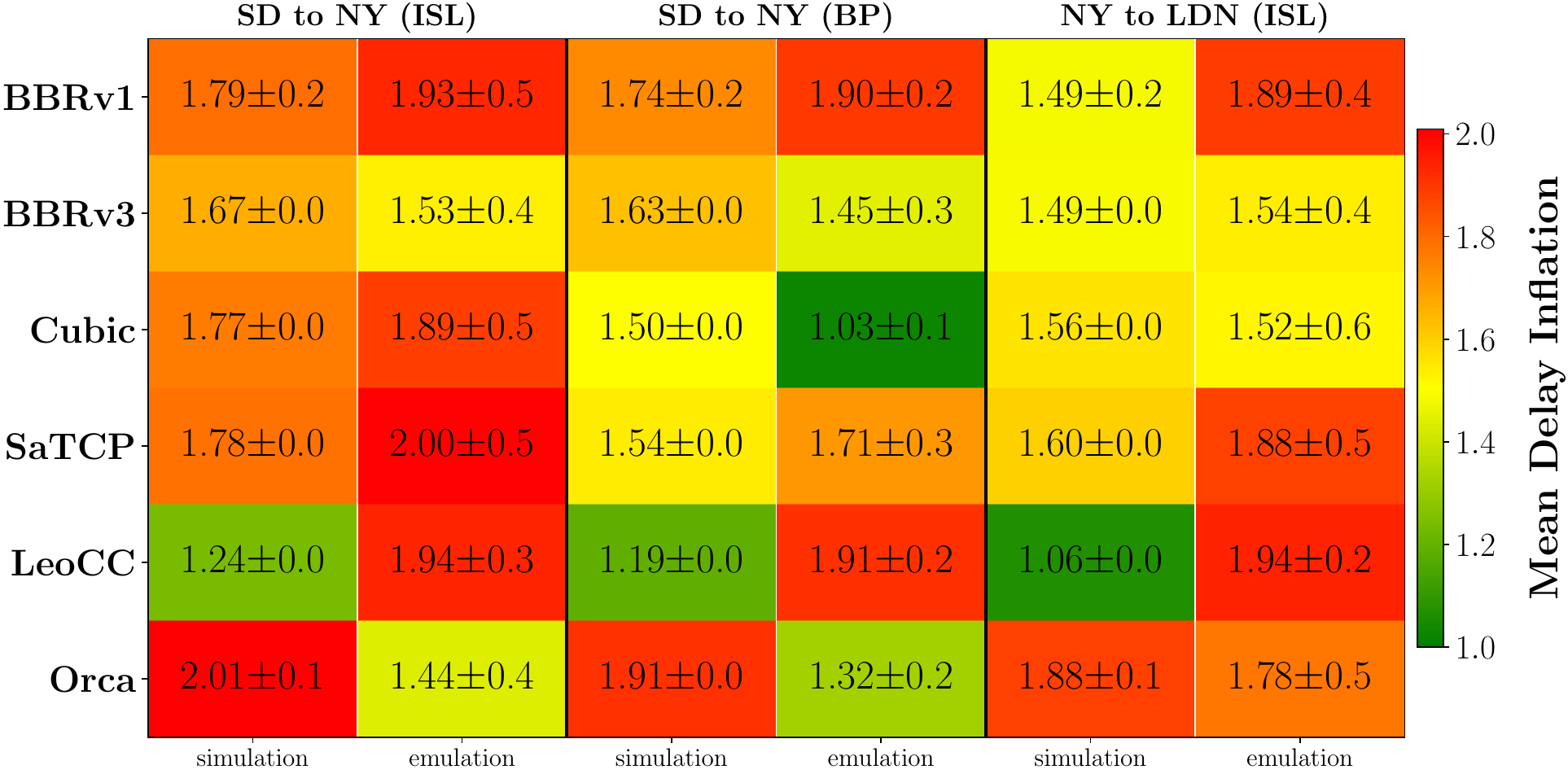}
        \caption{Delay Inflation}
        \label{fig:delay_heatmap}
    \end{subfigure}

    \caption{Performance of two competing flows experiencing identical base RTTs across various simulated/emulated LEO paths.}
    \label{fig:heatmap}
\end{figure*}

Having established that the extended simulator can reproduce the behaviour of a production transport stack under single-flow LEO-like dynamics, we next examine whether this carries forward into a multi-flow setup. Figure \ref{fig:heatmap} compares two competing flows with identical base RTTs across several LEO paths. Across most schemes, simulation and emulation preserve broadly similar fairness trends, with the exception of Orca. The discrepancy in Orca is likely caused by differences in the training environment and RL implementation: our retrained Orca uses soft actor-critic, whereas the original implementation uses twin delayed deep deterministic policy gradient (TD3)~\cite{orca}, due to RayNet constraints. We see similar trends in terms of average delay inflation (shown in Figure \ref{fig:delay_heatmap}). The higher delay inflation observed for LeoCC reflects its sensitivity to reconfiguration timing: even small differences in how simulation and emulation realise handover duration can change how long LeoCC preserves or refreshes its path model.

\begin{figure*}[t]
    \centering

    \begin{subfigure}[t]{0.48\textwidth}
        \centering
        \includegraphics[width=\linewidth]{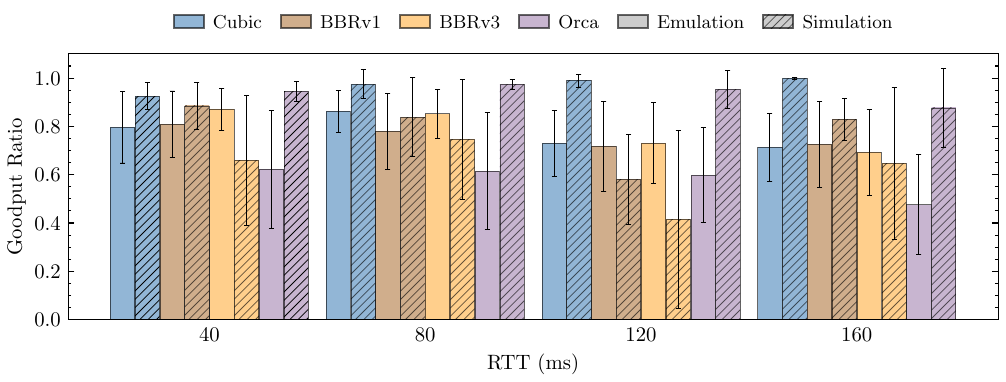}
        \caption{Both flows experience the same base RTT shown on the X-axis.}
        \label{fig:intra_rtt}
    \end{subfigure}
    \hfill
    \begin{subfigure}[t]{0.48\textwidth}
        \centering
        \includegraphics[width=\linewidth]{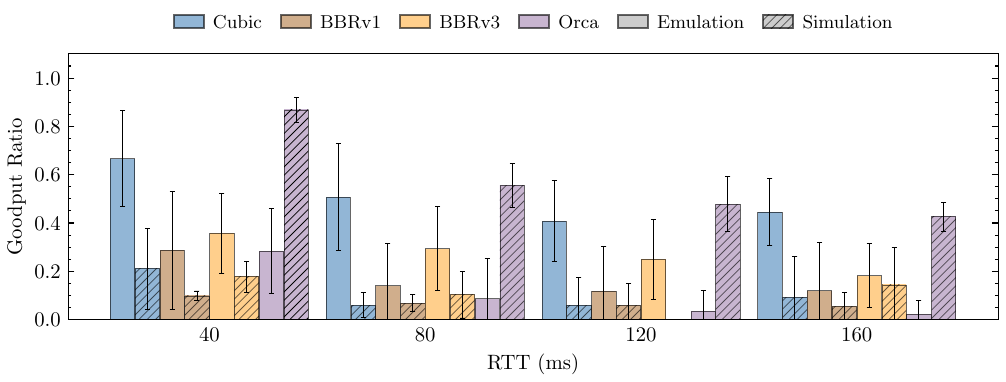}
        \caption{Starting flow has a base RTT of 20ms, the joining flow's base RTT is shown on the X-axis.}
        \label{fig:inter_rtt}
    \end{subfigure}

    \caption{Goodput ratio of two competing flows experiencing (a) identical RTTs. (b) different RTTs.}
    \label{fig:fairness_analysis}
\end{figure*}

\subsection{Fairness analysis under RTT asymmetry}

Fairness studies in LEO networks require comparing flows that traverse different satellite paths, for example sender-receiver pairs located in different geographic regions and therefore experiencing different RTT dynamics. This is difficult to study in current emulation frameworks, which cannot realistically instantiate a full constellation and multiple geographically distinct endpoint pairs at scale. Instead, emulators typically reproduce one or a small number of controlled path profiles, which limits their ability to expose inter-RTT fairness effects that emerge from constellation-wide path diversity.

Inter-RTT fairness is a particularly challenging regime in which traditional CC schemes often underperform. To date, no general heuristic has been shown to reliably guarantee fairness across flows with heterogeneous RTTs. This regime is particularly challenging for BBRv1, as the flow experiencing the higher RTT will hold more packets in flight, as it observed a larger BDP, claiming more buffer occupancy than the flow experiencing the lower RTT. In the case of BBRv3, its drain parameters have been tuned to increase its inter-RTT fairness \cite{bbrv3}, which is also reflected in simulation, where BBRv3 achieves a higher goodput ratio compared to BBRv1. On the other hand, the performance of Orca is quite different in simulations compared to emulations, showing much higher fairness, much like in the previous section. More broadly, the result reinforces that RL-based CC is harder to replicate exactly across environments than conventional schemes. This is especially true when certain behaviours, like fairness in this case, are not explicitly included in its state or reward formulation.

\subsection{Reproducibility vs Replicability}


Simulation offers strong reproducibility. Identical seeds yield repeatable results across runs and host environments, because the entire stack is a deterministic model advanced in virtual time. This is ideal for regression testing and for isolating the effect of a single parameter, but reproducibility does not imply replicability. A simulation can reproduce the same trace exactly while still failing to replicate the production implementation. This is particularly visible for BBRv3, where behaviour depends on implementation details beyond the CC algorithm. Linux rate samples are shaped by ACK timing, pacing, send-buffer state, socket accounting, and loss recovery. These assumptions do not map directly onto a discrete-event simulator. Implementation details also matter: Linux BBR uses fixed-point integer arithmetic for several rate and gain calculations, whereas simulator implementations often use floating-point values or abstract event-time measurements. Small differences in sampling, rounding, or update timing can accumulate into different pacing rates, congestion-window updates, and recovery decisions.

Emulation preserves the implementation behaviour that simulation abstracts away, since it runs the unmodified kernel stack on real hardware in wall-clock time, but that stack becomes a source of uncontrolled timing noise. Even where the queueing disciplines and pacing are scheduled at fine resolution, when each timed event is actually serviced varies with the state of the host. Identical configurations thus yield different packet arrivals, RTT samples, and delivery-rate estimates, and BBR diverges accordingly. Repeatability is statistical rather than exact: we repeat each configuration multiple times and report aggregated metrics with their variance rather than any single trace.

For RL-based CC, reproducibility depends on the full learning pipeline, not only on the simulator. A deterministic simulator can replay a fixed learned controller when the controller, random-number generators, and execution path are all fixed. However, reproducing the controller obtained through training is a stronger requirement. Even with fixed random seeds, training may diverge because floating-point operations, parallel execution, hardware backends, library versions, and update ordering can introduce small numerical differences. These differences can slightly alter observations, rewards, gradients, or policy updates, and may compound over training into a different learned controller. Simulation can therefore reproduce the network environment, but reproducing the learned controller requires controlling the complete training process.


\begin{figure}
    \centering
    \includegraphics[width=\columnwidth]{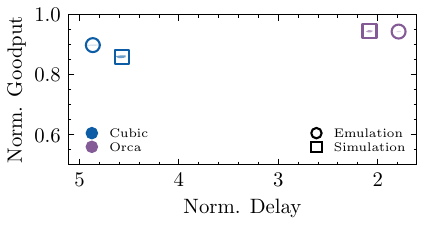}
    \vspace{-0.5cm}
    \caption{Aggregate normalised network goodput and normalised delay for two competing flows.}
    \label{fig:RLEfficiency}
    \vspace{-0.4cm}
\end{figure}

\subsection{Evaluation of Reinforcement Learning Congestion Control}
Learning-based CC presents difficulties that neither environment captures alone. RL schemes demand a massive and diverse body of training experience, and simulation is attractive precisely because its scalability supplies this cheaply and faster than real time. The risk is that simulation also overestimates robustness, since training and evaluation environments are often closely aligned. Training in simulation and evaluating under emulation is therefore the natural split, but only once the major discrepancies between the two are controlled. Two such discrepancies matter here.

RL agents tend to value large congestion windows early in training. Inflating the congestion window yields immediate reward and floods the path with excess packets. The two environments handle this differently: the surplus inflates event-queue and memory pressure in simulation, degrading throughput and occasionally causes crashes, whereas in emulation it merely drives the sending rate to hardware limits. Identical policy behaviour thus appears as a tooling failure in one environment and as physical saturation in the other.

Discrete-event simulators stop simulation time during each agent action step, distorting evaluation in two ways. Since the simulation clock does not advance during inference, even an oversized, slow model incurs no penalty, and simulation overstates its performance using real compute. The pause also means that there is zero delay between observing the network and acting on it. A delta is always present on a real stack and implicitly encoded in its state space which materially shapes the learned policy. A future eventuality of a simulation-trained model deployed on a real stack, would encounter this delay for the first time at inference, acting on observations that are out of synch relative to the behaviour it is controlling.

To assess the faithfulness of our Orca implementation, we evaluate its efficiency against Cubic in a two-flow setting. This comparison is important because Orca is not a clean-slate scheme, as it builds on Cubic behaviour and uses the learned policy to adjust its congestion window decisions. A single-flow experiment would provide only limited validation, since it cannot show how these adjustments behave under competing traffic. We therefore consider two concurrent flows, where correctness is reflected in aggregate goodput and delay rather than isolated per-flow behaviour. We plot the aggregate goodput and the average normalised delay in Figure \ref{fig:RLEfficiency} of a 100Mbps bottleneck bandwidth, 20ms base RTT and a 5 BDP buffer. In this scenario, the retrained simulated Orca performs similarly to the emulated implementation in terms of efficiency. The reward formulation of Orca includes a small delay budget that should allow it to explore bandwidth. This suggests that the implementation captures the intended efficiency trade-off in Orca’s reward design, where the policy is allowed to tolerate additional delay in order to achieve higher aggregate goodput.

\section{Conclusion}
\label{sect:conclusion}
The evaluation of modern CC, particularly in dynamic environments such as LEO satellite networks, can no longer rely on a single methodology. Accurate, deployment-relevant insight emerges only when scalable simulation, high-fidelity transport modelling, and real-stack emulation are combined in a principled workflow.

Simulation is essential for scale, control, and exploration of the design space, whilst emulation is essential for realism and validation, capturing transport-stack behaviour and implementation artefacts. Neither approach is sufficient alone; rigorous congestion control evaluation requires a deliberate combination of both. In particular, congestion control behaviour depends as much on supporting TCP mechanisms e.g. pacing, SACK, RACK, ACK timing,  as on the algorithm itself. Missing these mechanisms in simulation can produce qualitatively incorrect behaviour. Accurate evaluation demands modelling the full transport stack, not just the CC logic.

A key outcome of this work is that simulation frameworks must track deployed transport-stack behaviour if their results are to remain useful for CC research. Missing support for mechanisms such as pacing, SACK, RACK, ACK timing, and rate sampling can produce qualitatively incorrect behaviour, even when the high-level CC algorithm is implemented correctly.

Our experience has shown that LEO networks expose weaknesses in traditional evaluation methods, because they are stress tests for methodology.  Non-congestive loss, handovers, and RTT variability break many of the assumptions in existing CC and the dynamic topology magnifies small modelling inaccuracies into major behavioural differences.  Because  LEO networks amplify the discrepancies between simulation and reality, methodological rigour becomes critical.

Finally, reinforcement learning introduces new methodological challenges.  RL-based CC schemes, such as Orca, change the evaluation problem:
\begin{itemize}
\item Training depends heavily on simulation scale and speed.
\item Learned policies are sensitive to environment mismatch e.g., timing, delayed actions, and noise.
\item Simulation can overestimate robustness and performance, emphasising the need for testing and emulation.
\end{itemize}
As CC algorithms evolve, the research community needs hybrid workflows that preserve fidelity between simulated and emulated environments, so that deployment-facing conclusions can be drawn before widespread deployment.







\bibliographystyle{IEEEtran}
\bibliography{references.bib}

\end{document}